\documentclass[11pt, hyphens]{kimreview}
\usepackage{bm}
\usepackage{amsmath}
\usepackage{amssymb}
\usepackage{epstopdf} 
\usepackage{url}
\usepackage{xpatch}
\usepackage{dcolumn}
\usepackage{caption}
\newcolumntype{d}[1]{D{.}{.}{#1}} 
\usepackage{wrapfig}

\usepackage[style=numeric-comp,sorting=none,backend=biber, maxbibnames=50,giveninits=true]{biblatex}

\usepackage{hyperref}
\usepackage{xurl}
\hypersetup{breaklinks=true}
\renewbibmacro{in:}{}
\DeclareBibliographyAlias{article}{custom}
\DeclareFieldFormat{pages}{#1}
\DeclareFieldFormat{url}{\url{#1}}
\DeclareBibliographyDriver{article}{%
  \usebibmacro{bibindex}%
  \usebibmacro{begentry}%
  \usebibmacro{author/editor+others/translator+others}%
  \setunit{\labelnamepunct}\newblock
  \usebibmacro{title}%
  \newunit\newblock
  \usebibmacro{byauthor}%
  \newunit\newblock
  \usebibmacro{byeditor+others}%
  \newunit\newblock
  \printfield{version}%
  \newunit\newblock
  \printfield{journaltitle}%
  \newunit\newblock
  \iffieldundef{volume}
    {}
    {\printfield{volume}}
  \iffieldundef{number}
    {}
    {\printtext{(\printfield{number})}}
  \iffieldundef{pages}
    {}
    {\printtext{:\printfield{pages}},}
  \newunit\newblock
  \printfield{year}
  \newunit\newblock
  \iftoggle{bbx:eprint}
    {\usebibmacro{eprint}}
    {}%
  {\printtext{\printfield{url}}}
  \usebibmacro{finentry}%
}

\DeclareBibliographyDriver{misc}{%
  \usebibmacro{bibindex}%
  \usebibmacro{begentry}%
  \usebibmacro{author/editor+others/translator+others}%
  \setunit{\labelnamepunct}\newblock
  \usebibmacro{title}%
  \newunit\newblock
  \usebibmacro{byauthor}%
  \newunit\newblock
  \printfield{version}%
  \newunit\newblock
  \newunit\newblock
  \printfield{journaltitle}%
  \newunit\newblock
  \iffieldundef{volume}
    {}
    {\printfield{volume}}
  \iffieldundef{number}
    {}
    {\printtext{(\printfield{number})}}
  \newunit\newblock
  \iffieldundef{pages}
    {}
    {\printtext{:\printfield{pages},}}
  \newunit\newblock
  \newunit\newblock
  \iftoggle{bbx:eprint}
    {\usebibmacro{eprint}}
    {}%
  \newunit\newblock
  \usebibmacro{url+urldate}%
  \newunit\newblock%
  \usebibmacro{addendum+pubstate}%
  \setunit{\bibpagerefpunct}\newblock
  \usebibmacro{pageref}%
  \newunit\newblock
  \iftoggle{bbx:related}
    {\usebibmacro{related:init}%
     \usebibmacro{related}}
    {}%
  \usebibmacro{finentry}%
}

\DeclareFieldFormat{shorthandwidth}{#1}
\DeclareRangeChars{-}
\DeclareRangeCommands{\bibrangedash}

\DeclareCiteCommand{\cite}[\mkbibbrackets]
  {\usebibmacro{cite:init}%
   \usebibmacro{prenote}}
  {\usebibmacro{citeindex}%
   \usebibmacro{cite:comp}}
  {}
  {\usebibmacro{cite:dump}%
   \usebibmacro{postnote}}

\begin{document}


\title{The transformative capability of quantum-accurate machine learning interatomic potentials}


\authorOne{Alfredo A. Correa}
\affiliationOne{Quantum Simulations Group, Physics Division, Lawrence Livermore National Laboratory, Livermore, CA 94550, USA}

\authorTwo{Sebastien Hamel}
\affiliationTwo{Quantum Simulations Group, Physics Division, Lawrence Livermore National Laboratory, Livermore, CA 94550, USA}


\publishyear{2025}
\volumenumber{3}
\articlenumber{01}
\submitdate{May 16, 2025}
\publishdate{May 28, 2025}
\doiindex{10.25950/ce4db4e8}
\doilink{10.25950/ce4db4e8}

\paperReviewed
{Extreme Metastability of Diamond and its Transformation to the BC8 Post-Diamond Phase of Carbon}
{K. Nguyen-Cong, J.~T. Willman, J.~M. Gonzalez, A.~S. Williams,
A.~B. Belonoshko, S~G. Moore, A.~P. Thompson, M.~A. Wood,
J.~H. Eggert, M. Millot, L.~A. Zepeda-Ruiz, and I.~I. Oleynik}
{\href{https://doi.org/10.1021/acs.jpclett.3c03044}{J. Phys. Chem. Lett., 15(4):1152–1160 (2024)}}

\maketitle



\begin{abstract}
Many materials' properties and phase boundaries are generally not well known under extreme pressure and temperature conditions.
This is a consequence of the scarcity of experimental information and the difficulty of extrapolating approximations to the atomic interactions in such conditions.
Nguyen-Cong and colleagues,
in their publication (J.Phys.Chem.Lett.~\textbf{15}, 1152 (2024))~\cite{NguyenCong2024},
achieved an impressive result using a SNAP (Spectral Neighbor Analysis Potential), an interatomic potential for carbon obtained by machine learning techniques.
In a way, their contribution closes a full circle of research that spanned more than three decades.
\end{abstract}
\medskip


\section*{Full circle}

At the turn of the century, the scientific community was uncertain about the phase diagram of carbon, among other elemental systems.
There was an intriguing contrast among the group XIV elements of the periodic table that adopt a diamond structure.
Silicon and germanium both exhibited decreasing melting lines (from ambient pressure to \(\sim 11~\mathrm{GPa}\) for Si and \(\sim 9~\mathrm{GPa}\) for Ge).
At the same time, experiments on carbon showed an increasing melting line in the same pressure range.
\emph{Would the melting line of diamond keep increasing as higher pressures could be reached by experiments? Or would it have a maximum melting temperature?}

\begin{figure}[h!t]
    \centering
    \includegraphics[width=0.6\linewidth]{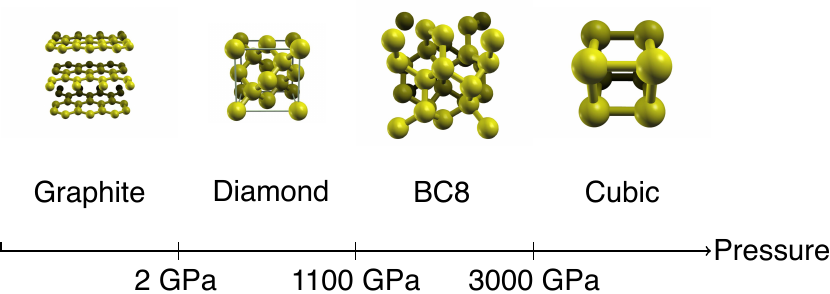}\\
    \caption{
    \emph{Solid crystalline phases of elemental carbon} and pressure conditions at which they are stable.
    In the diamond structure each atom is surrounded by 4 neighbors.
    BC8 is a body-centered structure with 8 atoms in the unit cell, local coordination is 4 but it has different global connectivity than diamond and is slightly more closed-packed.
    Graphite would have a coordination of 3, and cubic a coordination of 6.
    }
    \label{fig:carbon_coordination}
\end{figure}

\begin{figure}[!t]
    \centering
    \includegraphics[width=0.6\linewidth]{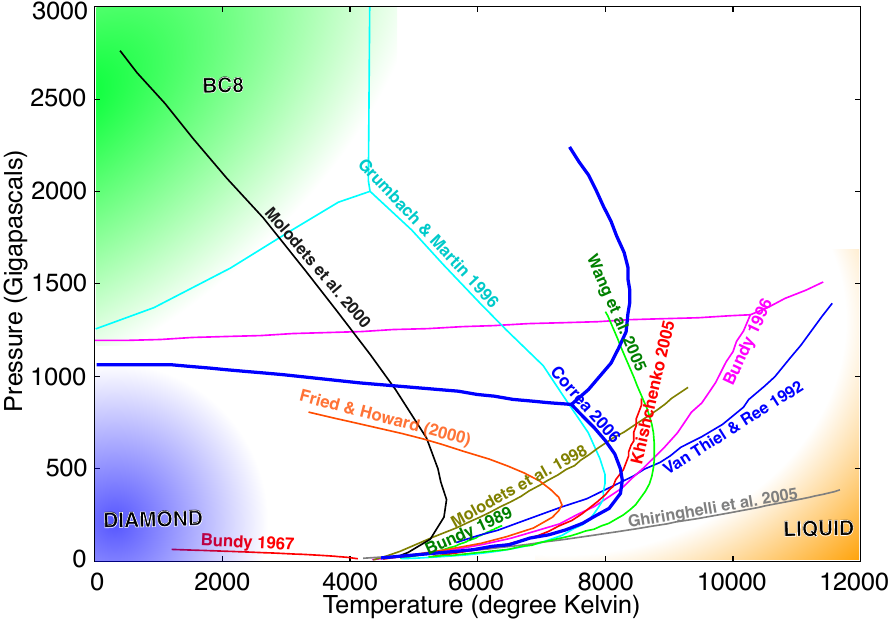}
    \caption{
\emph{Which carbon phase diagram would you choose? -- There was a significant lack of understanding regarding the carbon phase diagram at the turn of the century.}
The lines in the diagram represent different predictions for melting or transition points between various phases. These include the liquid phase (orange gradient region on the right), diamond (blue gradient on bottom left), and, in some models, the BC8 (or some unknown closed-packed structure) phase (green gradient on top left).
(White background describes unknown regions at the time.)
(The graphite phase, only stable at low pressures, is not displayed on this scale.)
Horizontal curves near \(1200~\mathrm{GPa}\) indicate the transition between diamond and BC8, showing the pressures and temperatures at which both structures are equally stable.
Some models feature melting lines that exhibit a maximum diamond melting temperature.
Later on, Nguyen-Cong \emph{et al.}~\cite{NguyenCong2024} obtained, using a machine learned SNAP potential, a phase diagram consistent with Correa \emph{et al.}~\cite{Correa2006} and explored the metastability of diamond, near and above the triple point (at \(T \simeq 7500~\mathrm{K}, P \simeq 900~\mathrm{GPa}\)).
(Graphic adapted from Ref.~\cite{CorreaThesis} and references therein.)
    }
    \label{fig:historic-models}
\end{figure}

\begin{figure}[!b]
    \centering
    \includegraphics[width=0.5\linewidth]{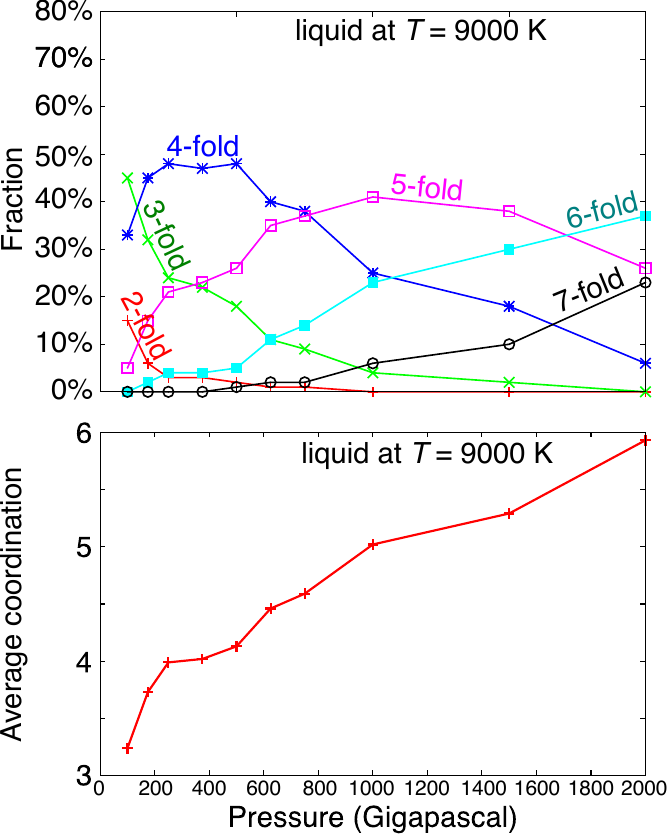}
    \caption{
    \emph{Liquid elemental carbon is capable of sustaining a variety of local atomic coordinations},
    posing a challenge for model interatomic potentials and as illustrated here by ab initio molecular dynamics (AIMD) results~\cite{CorreaThesis, Correa2006}.
    (top panel) Fraction of 2, 3, 4, 5, 6 and 7-fold coordinated atoms (number of first neighbors) as a function of pressure at fixed temperature (\(9000~\mathrm{K}\)) in \emph{liquid} carbon obtained from AIMD~\cite{CorreaThesis}.
    (bottom panel) Average coordination in the liquid phase as a function of pressure.
    At \(400~\mathrm{GPa}\) most atoms (\(\sim 50\%\) are found to have 4 first neighbors, reminiscent of the diamond solid structure, but other local coordination configurations coexist at all conditions).
    }
    \label{fig:carbon_coordination_vs_pressure}
\end{figure}

A plethora of effective interatomic potentials and crude estimations based on the Lindemann melting criterion predicted a whole array of results stressing the limits of phenomenological and extrapolated models (see Fig.~\ref{fig:historic-models});
especially in a range of conditions where the atomic coordination would change dramatically.
Across solid crystalline phases, these coordination changes occur in discrete steps (Fig.~\ref{fig:carbon_coordination}), but in liquids, the change is continuous and different local coordination can coexist, posing additional challenges for theoretical models (Fig.~\ref{fig:carbon_coordination_vs_pressure}).

It was only in 1996, with the advent of ab initio molecular dynamics (AIMD) results, that Grumbach and Martin~\cite{Grumbach1996} inferred that a decreasing melting line would occur for the diamond structure at much higher pressures (\(\sim 600~\textrm{GPa}\)).
These AIMD simulations showed a significant change in the coordination number of atoms in liquid carbon, while available semi-empirical or classical interatomic potentials used in molecular dynamics simulations could barely handle the different carbon atomic coordination (e.g., 3 or 4) as carbon compressed, let alone continuous or significant coordination as in the case of the liquid melt from AIMD.
Semi-empirical potentials, such as REBO and AIREBO derived from the work of Tersoff (1988)~\cite{Tersoff1988} had relatively simple forms and only a handful of adjustable parameters.
Molecular dynamics simulations using empirical potentials were not ready to give a complete picture of the phase-diagram or the Equation-of-State (EOS) calculations, which require a certain level of accuracy across a wide range of pressures and temperatures and phases of matter.
Such wide-ranging EOS are nowadays critical for designing and analyzing Inertial Confinement Fusion (ICF) and other dynamic compression experiments that depend on hydrodynamic simulations for analysis and experimental design.

\begin{figure}[!t]
    \centering
    \includegraphics[width=0.5\linewidth]{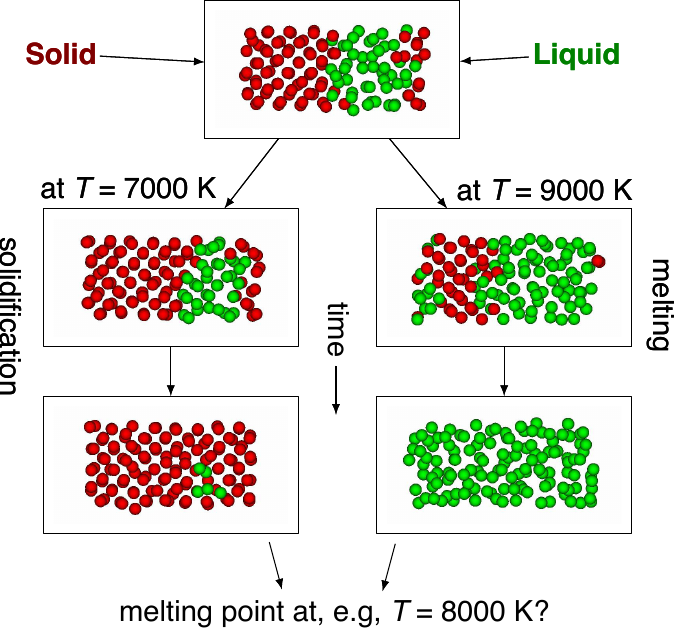}
    \caption{
    \emph{Direct simulation of the melting/crystallization process}.
    Snapshots illustrating the coexistence two-phase method with a total of 128
carbon atoms simulated by AIMD~\cite{CorreaThesis}.
The first row shows the initial condition where liquid and solid samples are put together, 64 atoms each.
The second row shows the progress during
a constant pressure-temperature simulation at two different conditions (\(7000~\mathrm{K}\) or \(9000~\mathrm{K}\)).
    The third row shows a final state at which either the solid or liquid takes over the simulation cell.
    After such outcome, we would infer that the melting point lies somewhere between the two simulated temperatures.
    We could try the midpoint next, a simulation at \(8000~\mathrm{K}\), to further bracket the melting point temperature.
    Carbon atoms are colored according to a local order identification of the solid (red) and liquid (green).
    We can appreciate the challenges defining the interfaces and criteria for melting and solidification in such small cells, a limitation imposed by the use of AIMD.
    \label{fig:two-phase}
    }
\end{figure}

The prevailing lack of accuracy of classical interatomic potentials motivated
some of us~\cite{Correa2006}
to turn to AIMD to study the EOS of high-pressure carbon
using two-phase coexistence simulations to bracket the melting line over a wide range of pressures.
In AIMD, the interatomic potential is calculated on the fly from expensive electronic structure (Density Functional Theory -- DFT) calculations without a specific functional form or fitting parameters, capturing in principle all the effects of chemical bonding (coordination effects) and metallic bonding (electron band energies).
The two-phase coexistence simulation method is based on the direct simulation of the relative stability of liquid and solid at well defined conditions (Fig.~\ref{fig:two-phase}).
The two-phase method is an alternative to methods that calculate the free-energies of independent phases that must match at the coexistence conditions.
The two-phase method is an illustration of a non-equilibrium simulation (the explicit time dynamics of an interface) to obtain an equilibrium property (the phase diagram).

The trade-off was that, given the computational cost of AIMD simulations, they could only simulate a few tens of atoms.
This meant that even if the interatomic potential had been perfectly accurate, it was harder to reach the thermodynamic (large \(N\)) limit.
This method could only afford to simulate \(N = 128\) atoms (\(64\) atoms for each phase).
With such a small number of atoms in each phase, the contribution of the interface versus the system's bulk behavior cannot be separated (all the atoms are, in a sense, part of the interface), which may be problematic for predicting bulk thermodynamic properties.

The two-phase coexistence approach was a relatively new proposed method applicable to simulations of comparable scale, spearheaded by Belonoshko~\cite{Belonoshko1994} (a coauthor on the commented article) and by others~\cite{Kubicki1992, Ogitsu2003}.

This same two-phase coexistence method is used by Nguyen-Cong \emph{et al.}~\cite{NguyenCong2024}, although now with a SNAP interatomic potential fitted to reproduce AIMD results and computationally cheap enough to enable simulations with millions of atoms so results closer to the thermodynamic limit would be assured.
They can also simulate true nucleation domains, which is relevant to defining the time-scale of the kinetics of phase transitions.

At the time, the AIMD results~\cite{Correa2006} were surprising because there were not many known examples of melting lines with negative slopes and they found it \emph{twice} for two different solid phases of carbon (diamond and BC8). 
The accurate location of these melting lines in a wide range of pressures \emph{pinned} a set of reference points for the free energy.
That is, once well-known models of solid free energy are established, we could obtain absolute values of entropy (!) of liquid carbon, which was then used to construct one of the first multiphase EOS based entirely on ab initio DFT~\cite{Correa2008}.
This new phase diagram and EOS were partially corroborated by the experimental work of Eggert (also a coauthor in the commented article) and colleagues~\cite {Eggert2009}.

The apparent success of ab initio molecular dynamics, along with the innovative two-phase method, likely left the community—and us—with mixed feelings.
While it demonstrated that classical interatomic potential models were still far from ideal, it also highlighted important advancements.

Indeed, ab initio methods proved effective for diagnosing equilibrium properties in elemental systems, validating decades of method development and showing predictive power under extreme temperature and pressure conditions.
However, questions lingered, and remained unanswered for at least another decade:
\emph{Would kinetics, along with systems featuring complex formulas, non-stoichiometric mixtures, and size effects, remain beyond the reach of these accurate potentials? Would size effects persist in our otherwise precise simulations indefinitely?}

\begin{figure}[ht]
    \centering
    \includegraphics[width=0.5\linewidth]{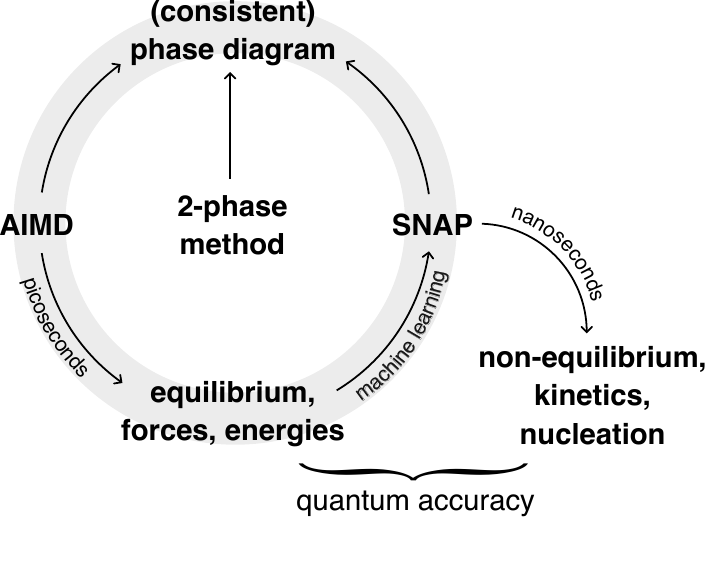}
    \caption{
    \emph{Full circle}. Ab initio molecular dynamics (AIMD) has proven an accurate tool for prediction of equilibrium properties and phase diagrams by means of picosecond-scale simulation in tens or few hundred atoms.
    With the advent of machine learning, this information can be used by complicated potential models such as SNAP, generated with machine learning and applied by Nguyen-Cong \emph{et al.}, that not only reproduce AIMD accuracy in the conditions of interest but also enables simulation of non-equilibrium, non-homogeneous systems, nucleation and kinetics in million-plus atom simulations, predicting the time scales required in the experiment to observe transitions between metastable phases.
    \label{fig:full_circle}
    }  
\end{figure}

Fast forward to today -- It is fascinating to observe how everything comes full circle.
Nguyen-Cong \emph{et al.}\ used a DFT-based machine-learned interatomic potential, specifically SNAP, along with millions of atoms, to generate a phase diagram that aligns well with the results from the small system size AIMD studies~\cite{Correa2006} (Fig.~\ref{fig:full_circle}). 

It is reassuring for many of us that the new phase diagram results of Nguyen-Cong \emph{et al.} with millions of atoms are surprisingly compatible with pure AIMD results on small cells;
showing that DFT's quantum accuracy can be efficiently transferred into a sufficiently complex interatomic potential, such as SNAP, to create phase diagrams across a wide range of densities and temperatures.
This approach effectively captures changes in atomic coordination while simultaneously addressing the behaviors of both insulating (solid) and metallic (solid and liquid) phases.

Of course, Nguyen-Cong \emph{et al.}\ results go much beyond as we will see; enabled by the simulation of millions of atoms with quantum accuracy, they were able to tackle questions sparked by a renewed interest in carbon at extreme conditions for BC8 synthesis, which is now, in principle, achievable by the double- or multi-shock compression pathway at the National Ignition Facility~\cite{Osolin2023}.

\section*{Renewed interest in carbon under extreme conditions}

Although the phase line for the diamond to BC8 transition is predicted by AIMD and SNAP-based MD to be around \(10~\mathrm{Mbar}\), the direct identification of a post-diamond phase (be it BC8 or another phase) through X-ray diffraction (XRD) experiments has not yet been successful, with the recent experiments of Lazicki \emph{et al.}~\cite{Lazicki2021} using quasi-isentropic (or ramp) compression reaching \(20~\mathrm{Mbar}\) but finding no evidence of a post-diamond phase transition.

Critically, the synthesis and characterization of BC8 would require a spontaneous nucleation of this phase in the time-scale of the experiment.
Nucleation is the process where a stable phase forms from a metastable phase;
microscopically, it is a more general case of the two-phase evolution of Fig.~\ref{fig:two-phase}, but it is also spontaneous and requires many more atoms for its simulation.

The work of Nguyen-Cong and collaborators presents an intriguing proposition: While the {\em equilibrium} phase boundary for the diamond-to-BC8 transition may well be at \(10~\mathrm{Mbar}\), the \emph{nucleation} of the BC8 phase on the few nanosecond timescale of dynamic compression experiments leads to a ``narrower'' pressure and temperature range where BC8 might be observed (between \(16-22~\mathrm{Mbar}\) and between \(4000-5000~ \mathrm{K}\)).
This is a key result, and such molecular dynamics simulations are playing an important role in the design of experiments for a new Discovery Science campaign on post-diamond phases of carbon under extreme conditions at the National Ignition Facility~\cite{Osolin2023}.

The development of quantum-accurate interatomic potentials that scale efficiently to several million or even billions of atoms on modern High-Performance Computing machines is a game changer in the study of material properties under extreme conditions.
This enabled, in turn, accessing both the nanosecond time-scale, which is on par with experiments, and the length-scale necessary for the simulation of nucleation with quantum (DFT) accuracy.
This is an impressive feat that was unimaginable 20 years ago using brute-force AIMD.

Machine learning interatomic potentials have proven highly versatile and have overcome the limitations of the simple-forms characteristic of early empirical potentials.
This overcame problems related to variable coordination (static and dynamic) and could model insulating and metallic systems in a variety of systems:
From capturing the coordination changes of liquid carbon, to modeling tantalum~\cite{Thompson2015}, to obtaining the correct anharmonicity of 
tungsten~\cite{Mo2024}, elements at opposite ends of the periodic table, correctly adjusted SNAP potentials have been shown to be flexible, and what we now describe as `quantum-accurate'.

\section*{Outlook for quantum-accurate potentials}

On the experimental side, extreme conditions of pressure and/or temperature are difficult to study in isolation, so instead we opt to contain the samples \emph{in time} using dynamic compression (inertial confinement) experiments, often on the time scale of nanoseconds for laser-based experiments.
We design and analyze these experiments in part by running hydrodynamic simulations that require the knowledge of reasonably accurate wide-range EOS models, transport, and constitutive properties (e.g., strength) of all materials involved.
These models are built using a combination of electronic structure calculations, molecular dynamics calculations, free energy ansatz, and prior experimental results.

In addition to helping with the EOS, the use of DFT-trained machine-learned interatomic potentials (MLIP) such as SNAP offers a practical way to consider the impact of more realistic systems, i.e. polycrystalline samples instead of perfect crystal, grain boundaries, and defects.
It enables calculation of properties of liquid carbon such as viscosity, direct shock simulations to be performed, and various mechanisms for plastic flow to be studied.
Already Nguyen-Cong \emph{et al.}~\cite{NguyenCong2024} are looking at the time scale of nucleation and growth and the kinetics of phase transitions.
Including such effects in hydrocodes is not yet routine, but it has been done for a liquid-solid transition in iron~\cite{Kraus2022} and a similar study would be interesting to perform for carbon and could help to better relate the result of dynamic compression experiments to equilibrium properties of diamond and BC8.

We note there are still certain limitations with MLIP models predicting energies and forces using only structural information.
When the temperature is sufficiently high, a significant fraction of the electrons are thermally excited.
In Mermin's formulation of finite-temperature DFT this effect is approximated using a Fermi distribution of the occupation numbers of Kohn-Sham orbitals and this electronic entropy has an impact on the forces~\cite{Wentzcovitch1992}.
Hence, it would be preferable if MLIP models were augmented with the contribution from the electronic entropy, which would allow the matching of DFT-calculated Hugoniots for much larger temperatures (instead of having a systematic shift as in Figure 2 in Ref.~\cite{Willman2022} which use the same carbon SNAP potential).
See, for example, Refs.~\cite{Zhang2020} and ~\cite{Kumar2024} for work along these lines, although we note that these particular approaches are restricted to equilibrium simulations.
A more general method that would be applicable to non-equilibrium situations has yet to be developed.

Nevertheless, with new science and predictions coming at a fast pace from these methods, it is difficult to overemphasize the transformative effects of quantum-accurate machine learning interatomic potentials, particularly in the field of matter at extreme conditions.

\section*{Acknowledgments}
Work performed under the auspices of the US Department of Energy by Lawrence Livermore National  Laboratory under contract DE-AC52-07NA27344.
S.H. acknowledges funding from the Laboratory Directed Research and Development (LDRD) Program at LLNL under project tracking code 23-SI-006.

%
%
\printbibliography

\end{document}